\documentclass[aps,twocolumn, amsmath,amssymb, superscriptaddress, reprint]{revtex4-1}
\usepackage[labelformat=simple]{subcaption}

\usepackage{xcolor}

\usepackage{hyperref}
\hypersetup{hypertexnames=false,
	bookmarksopen=true,%
	bookmarksnumbered=true,%
	colorlinks=true,%
    	linkcolor=blue,
    	citecolor=blue,
    	filecolor=blue,
    	urlcolor=blue,
    	linkbordercolor=blue,
    	citebordercolor=blue,
    	filebordercolor=blue,
    	urlbordercolor=blue,
	pdfborder={0 0 0.5},
	pdfstartview=FitH,%
	pdfnewwindow=true
}
\usepackage{amsmath,amsthm,amssymb}
\usepackage{graphicx}
\usepackage{siunitx}

\begin{document}

\title{Nanoparticle Classification in Wide-field Interferometric Microscopy by Supervised Learning from Model}

\author{Oguzhan Avci}
\thanks{Correspondence to: \href{mailto:oguzhan@bu.edu}{oguzhan@bu.edu}}
\affiliation{Department of Electrical and Computer Engineering, Boston University, Boston, MA 02215}
\author{Celalettin Yurdakul}
\affiliation{Department of Electrical and Computer Engineering, Boston University, Boston, MA 02215}
\author{M. Selim \"Unl\"u}
\affiliation{Department of Electrical and Computer Engineering, Boston University, Boston, MA 02215}
\affiliation{Department of Biomedical Engineering, Boston University, Boston, MA 02215}

\begin{abstract}
Interference enhanced wide-field nanoparticle imaging is a highly sensitive technique that has found numerous applications in labeled and label-free sub-diffraction-limited pathogen detection. It also provides unique opportunities for nanoparticle classification upon detection. More specifically, the nanoparticle defocus images result in a particle-specific response that can be of great utility for nanoparticle classification, particularly based on type and size. In this work, we combine a model based supervised learning algorithm with a wide-field common-path interferometric microscopy method to achieve accurate nanoparticle classification. We verify our classification schemes experimentally by using gold and polystyrene nanospheres.
\end{abstract}

\maketitle

\section{Introduction}

Interferometric nanoparticle imaging in a common-path configuration has gained significant attention for its ability to detect sub-diffraction-limited low-index biological nanoparticles, and its simple, cost-effective, and high-throughput setup \cite{AvciReview}. It enables highly sensitive detection of nanoscale particles by providing the means to enhance the signal through the interference between the scattered and reflected reference fields. To do so, it uses a layered sensor typically comprised of a thin layer of SiO$_2$ atop a Si substrate in a common-path interferometry configuration. It achieves enhanced scattering of nanoparticles in the light collection direction given the optimized thickness of this glass layer, similar to engineering a dipole antenna directivity \cite{AvciPhysicalModel,Avci:17,NovotnyBook}. Its signal is also affected by the polarizability of the particle, amplitude of the reference field and the phase lag between them as discussed in more detail in \cite{AvciPhysicalModel}.

\begin{figure}[h]
\begin{subfigure}{\linewidth}
\centering
\includegraphics[width=0.9\linewidth]{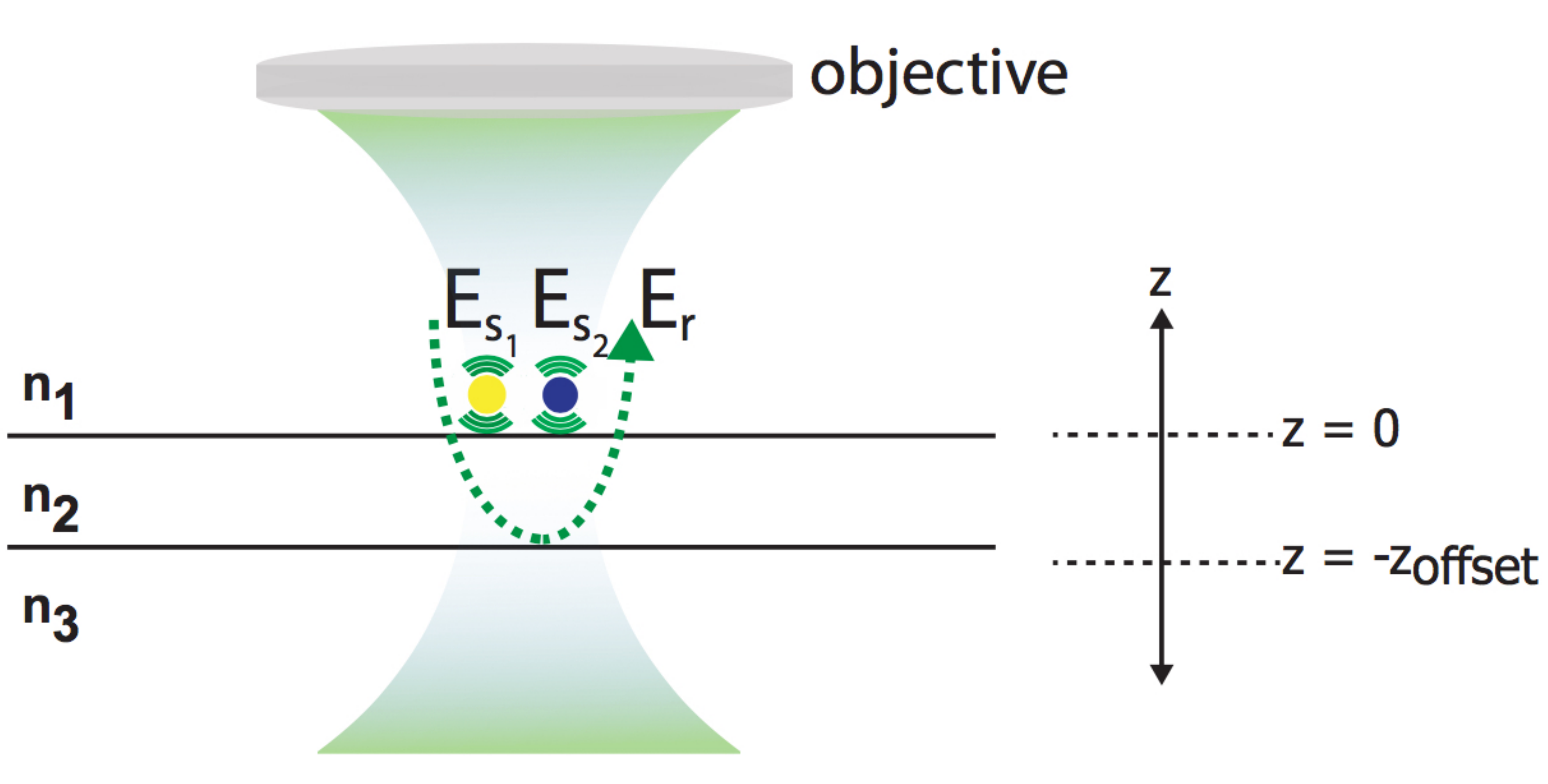} 
\captionsetup{justification=centering}
\caption{}
\label{fig:fig1a}
\end{subfigure}%

\begin{subfigure}{\linewidth}
\centering
\includegraphics[width=\linewidth]{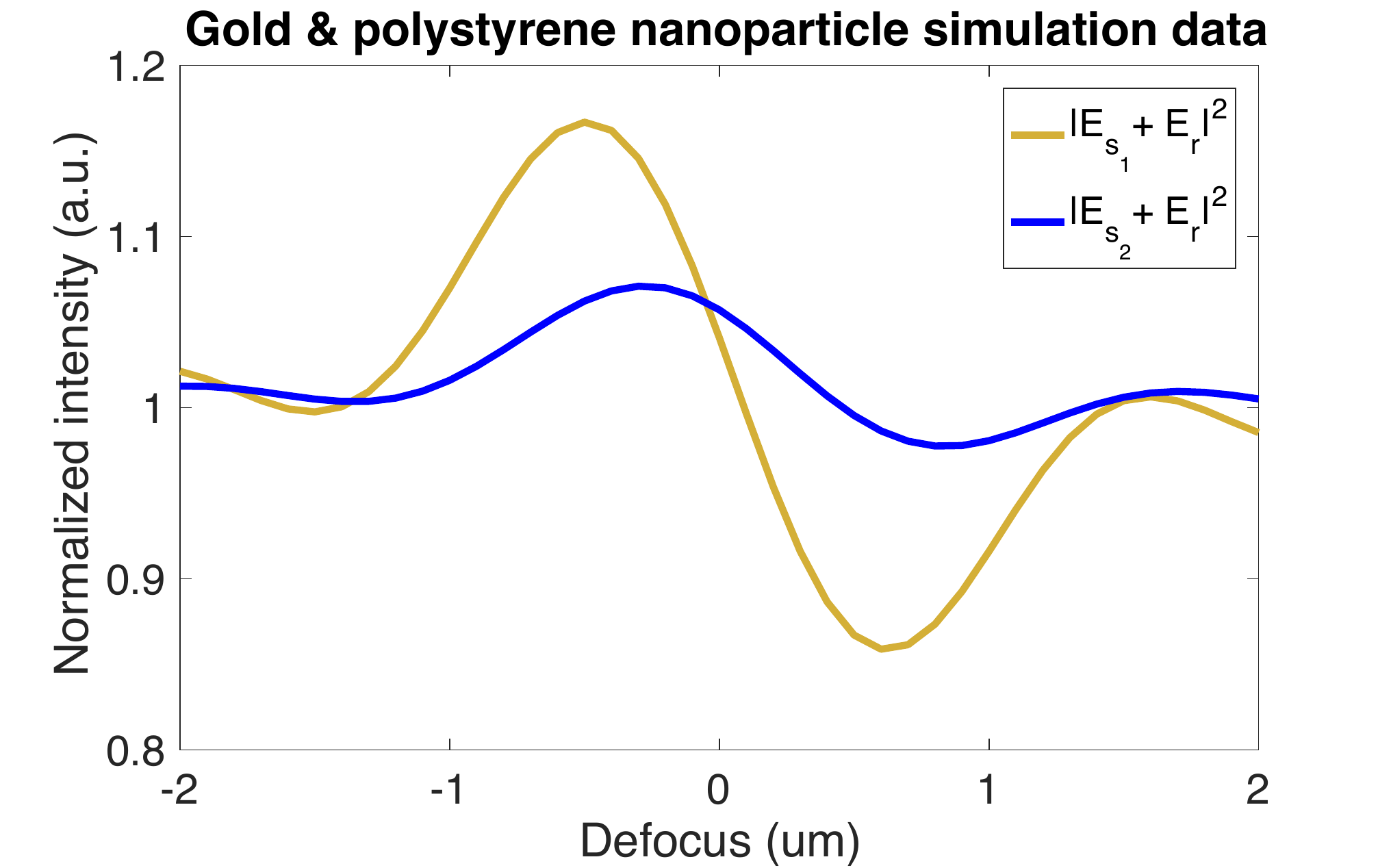}
\captionsetup{justification=centering}
\caption{}
\label{fig:fig1b}
\end{subfigure}%
\caption{The defocus in common-path wide-field interferometric nanoparticle imaging: (a) the schematic of the setup illustrating the case when sensor surface is in the same plane as the focal plane of the objective ($z = 0$) (adapted from \cite{AvciPhysicalModel}); the defocus takes place by moving this layered substrate in the axial direction (z), (b) the simulated interferometric signal for gold nanosphere in 30 nm radius and polystyrene nanosphere in 37 nm radius.}
\label{fig:1}
\end{figure}

The common-path interferometric nanoparticle imaging has been mainly demonstrated for label-free virus detection and sizing \cite{Daaboul:14,Scherr:16}. In these earlier studies, the particle detection and subsequent sizing relied on the intensity reading at a single focal plane. Under ideal circumstances, where the sample only contains a certain kind of nanoparticles (i.e., optical properties known a priori), this type of blind digital detection and sizing of nanoparticles based on a single focal plane image could yield reliable and repeatable results. However, polydispersity in nanoparticles and morphological variations on the sensor surface can lead to inaccurate detection and sizing in a single focal plane image. In such a case, a more robust detection and particle characterization is deemed necessary, and can be realized by defocus data stack acquisition strategy \cite{AvciPhysicalModel}. In fact, interferometric nanoparticle imaging can provide unique opportunities for the detection and visualization of weakly-scattering sub-diffraction-limited particles, as well as for their classification upon detection. Recently, with the development of a rigorous physical model \cite{AvciPhysicalModel}, it has been suggested that a more robust approach toward nanoparticle detection and discrimination can be achieved \cite{JSTQE}. In particular, the distinctive defocus signature of the interferometric signal can be of great utility to classify nanoparticles, as the additional phase introduced by the axial shift results in a particle-specific signature of nanoparticle response as shown in Fig. \ref{fig:fig1b} \cite{AvciPhysicalModel}. Here, the defocus is defined as the axial shift of the top sensor surface from the focal plane of the objective lens as illustrated in Fig. \ref{fig:fig1a}. 

A robust nanoparticle classification can especially be of great significance for the high-throughput interferometric biological-particle imaging where thousands of particles are detected simultaneously within a single field-of-view in a labeled and/or label-free fashion. For instance, when a multiplex detection of label-free viruses and antigens labeled with metallic particles take place on the same sensor surface \cite{Unlu:2015}, an accurate classification of the particles in the image becomes extremely important for accurate quantitative measurements. In another case, an accurate size discrimination of a polydisperse pathogens captured on the sensor surface can be of great importance in certain diagnostic applications. Therefore, it is clear that upon detecting the target nanoparticles with interferometric microscopy, there is a need for a reliable nanoparticle classification, particularly in terms of type and size. In the light of this need, here, we propose a way to classify interferometrically-detected nanoparticles based on size and type using experimentally obtained interferometric signal and the physical model detailed in \cite{AvciPhysicalModel}. The defocus signatures for nanoparticles of different types and sizes are utilized to realize a machine learning based nanoparticle classification scheme using support vector machine classifier (SVM). SVM classifier is a supervised learning algorithm that has been employed in various fields in the realm of image classification, ranging from face recognition to cell sorting \cite{imageclassification, facerecognition, humandetection, EMG, SVMcancertissue, SVMgeneselection}. SVM classifiers construct hyperplanes that linearly separate the labeled feature vectors in the training data set.

The training data set in this technique is provided by the simulated interferometric signals for different type and size nanoparticles. Once the supervised learning with simulated data is carried out, the experimental observations are fed into the algorithm where classification takes place in two steps: first the nanoparticle type is determined in terms of its dielectric characteristics, i.e., resonant (e.g., gold) or non-resonant (e.g., polystyrene), and secondly, given the type, the nanoparticle size is determined. We experimentally validate our technique by imaging a sample that has a mixture of gold and polystyrene nanospheres, and then classifying the detected nanoparticles in terms of type and size. Our method has a potential application in multiplexed interferometric microscopy experiments, where classification of nanoparticles can not only provide further information about the target but also eliminate the false count of spurious signal (noise) due to non-specific binding events.
\begin{figure*}[ht]
\begin{subfigure}{.4\linewidth}
\centering
\vspace*{-1in}\includegraphics[width=0.9\linewidth]{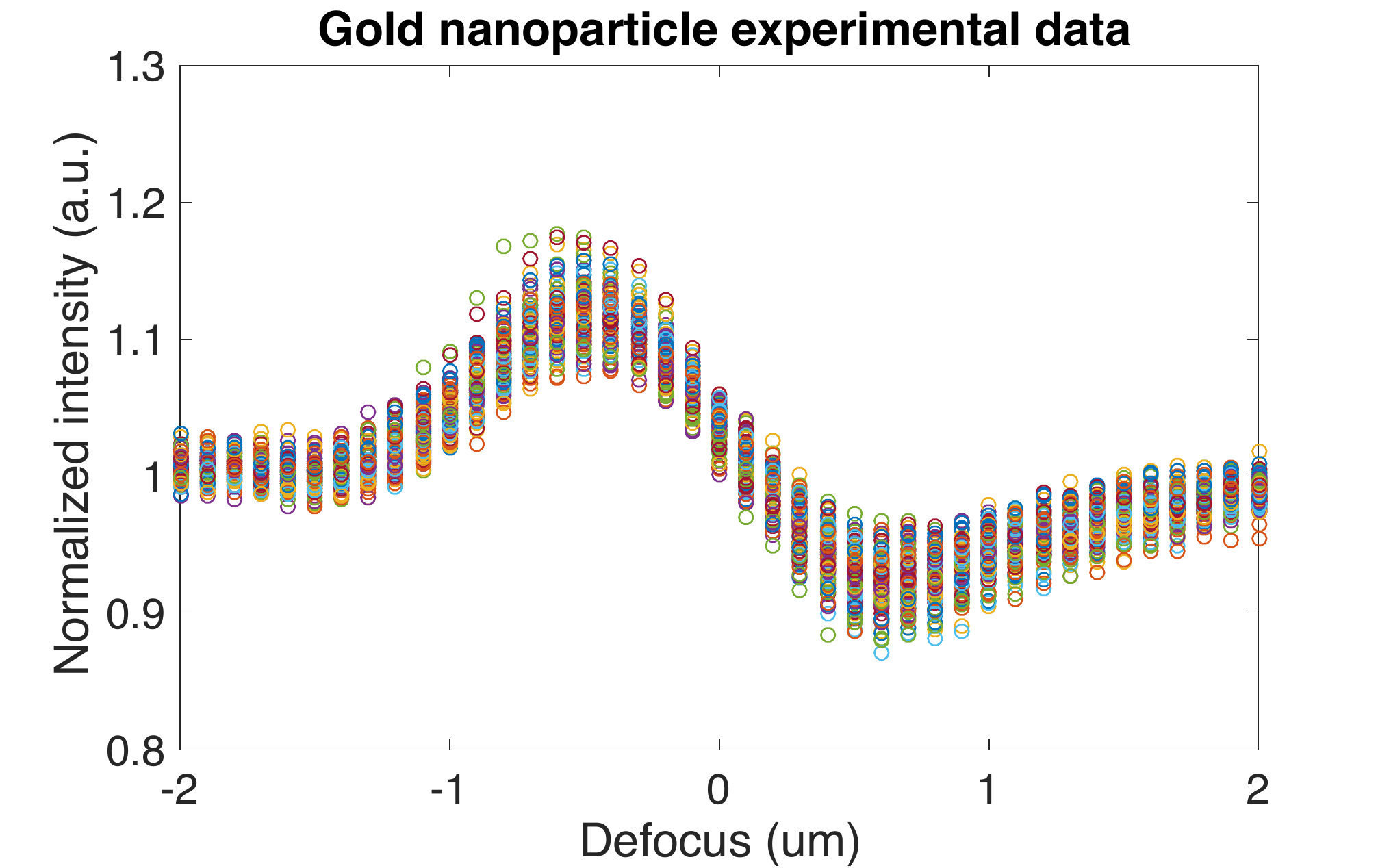} 
\captionsetup{justification=centering}
\caption{}
\label{fig:fig2a}
\end{subfigure}%
\begin{subfigure}{.2\linewidth}
\centering
\vspace*{1in}\includegraphics[width=\linewidth]{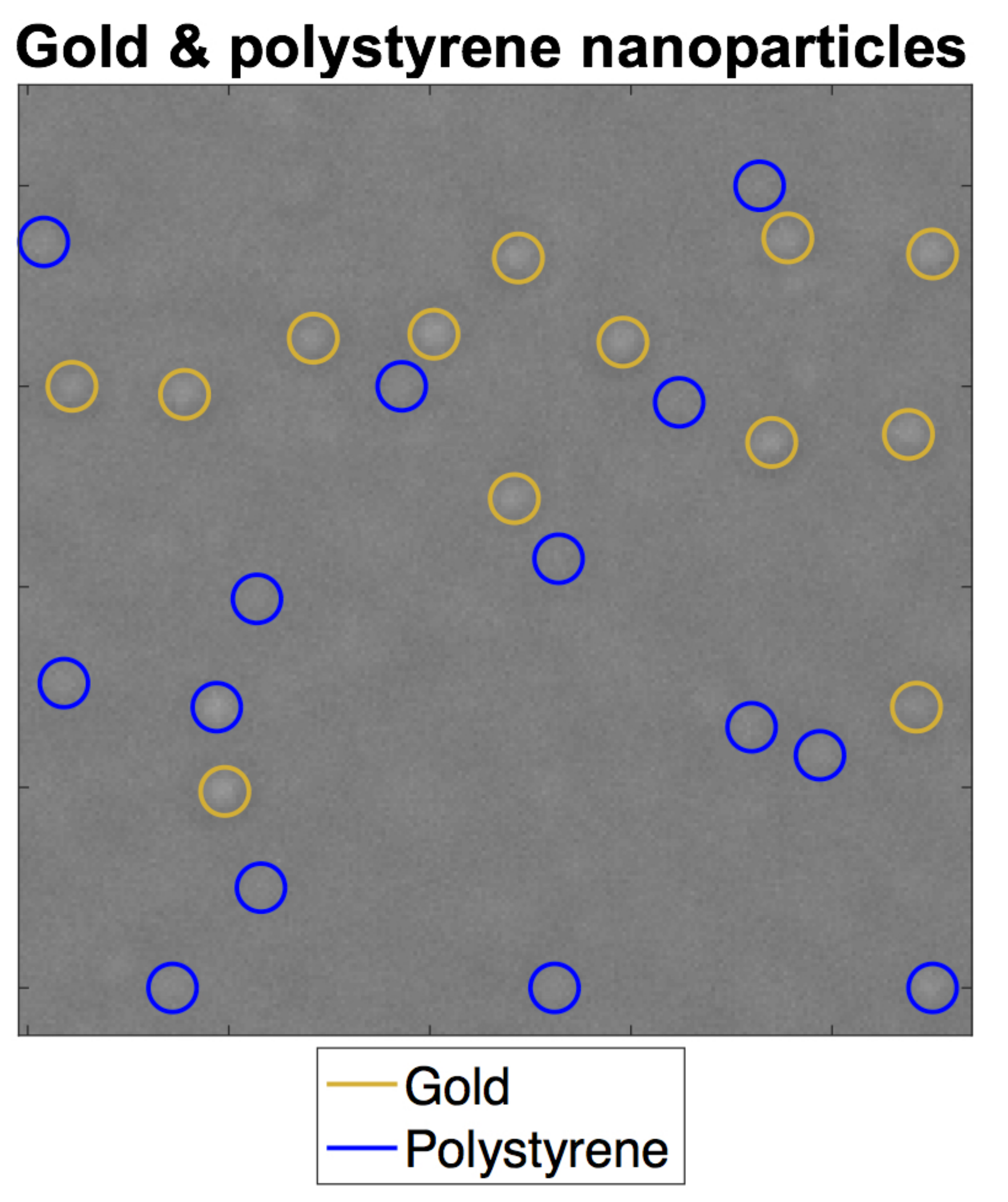} 
\captionsetup{justification=centering}
\caption{}
\label{fig:fig2b}
\end{subfigure}%
\begin{subfigure}{.4\linewidth}
\centering
\vspace*{-1in}\includegraphics[width=0.9\linewidth]{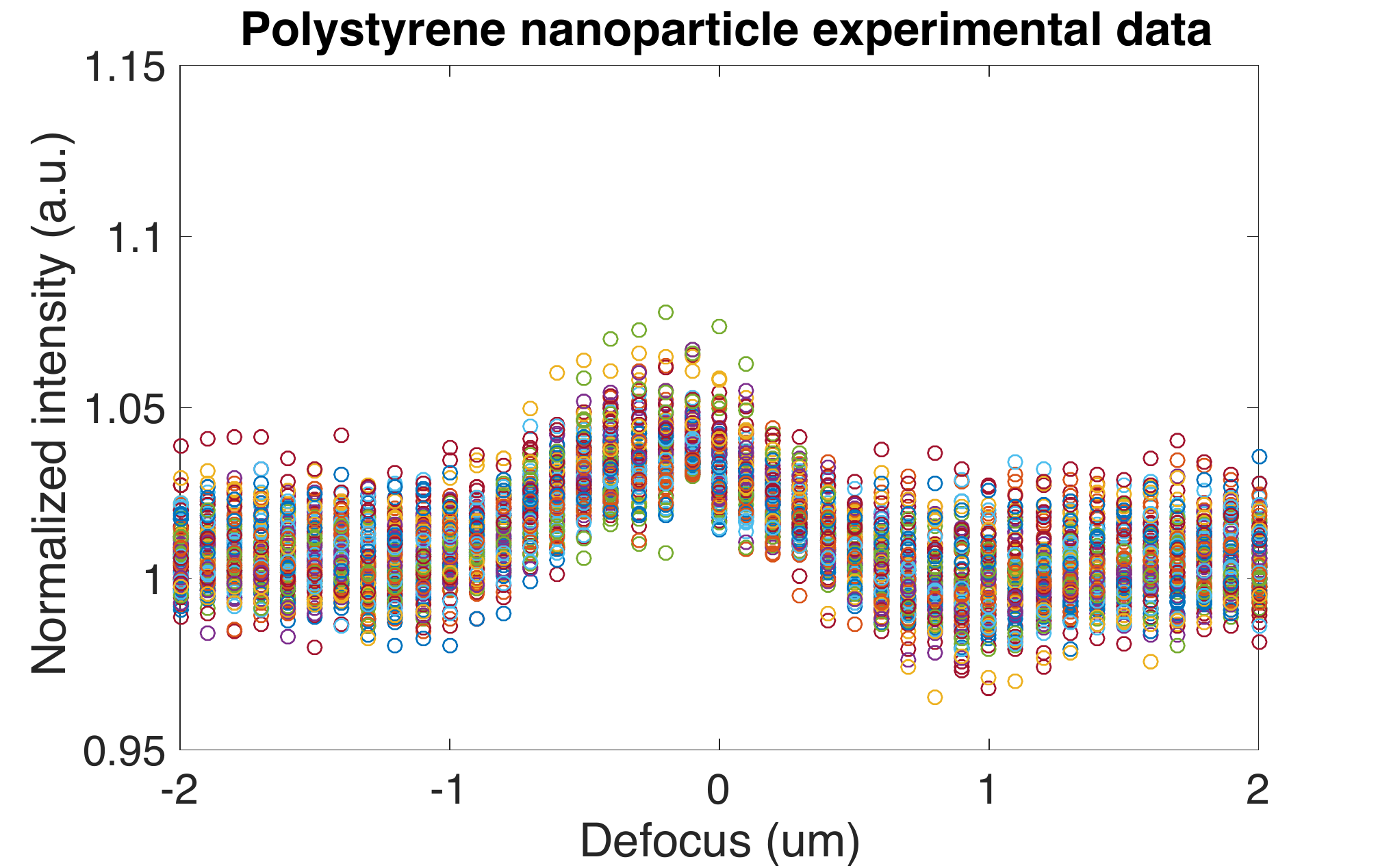} 
\captionsetup{justification=centering}
\caption{}
\label{fig:fig2c}
\end{subfigure}%

\begin{subfigure}{.4\linewidth}
\centering
\vspace*{-1in}\hspace*{-.5\linewidth}\includegraphics[width=.9\linewidth]{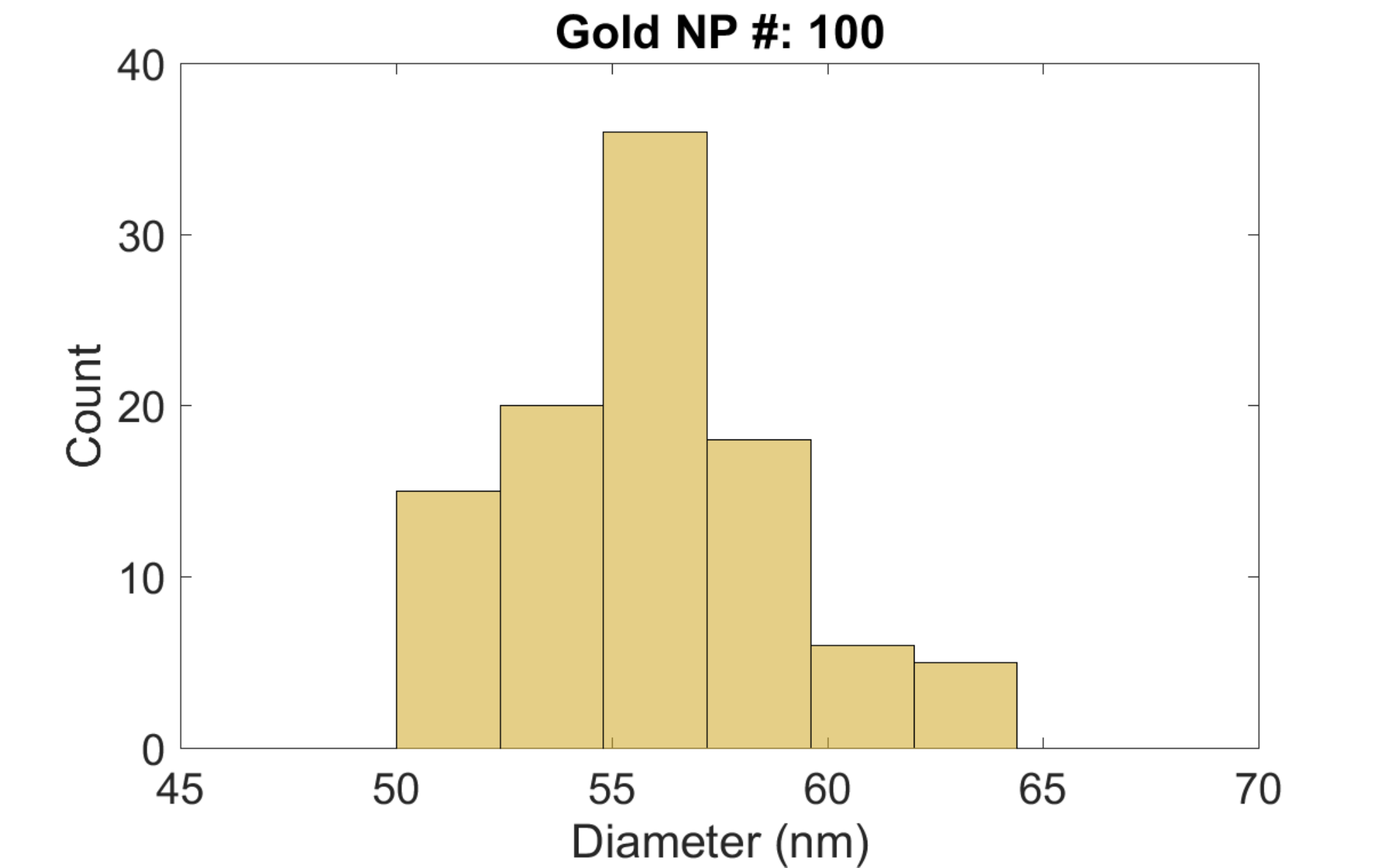} 
\captionsetup{justification=centering}
\captionsetup{oneside,margin={-.5\linewidth,0cm}}
\caption{}
\label{fig:fig2d}
\end{subfigure}%
\begin{subfigure}{.4\linewidth}
\centering
\vspace*{-1in}\hspace*{.3\linewidth}\includegraphics[width=.9\linewidth]{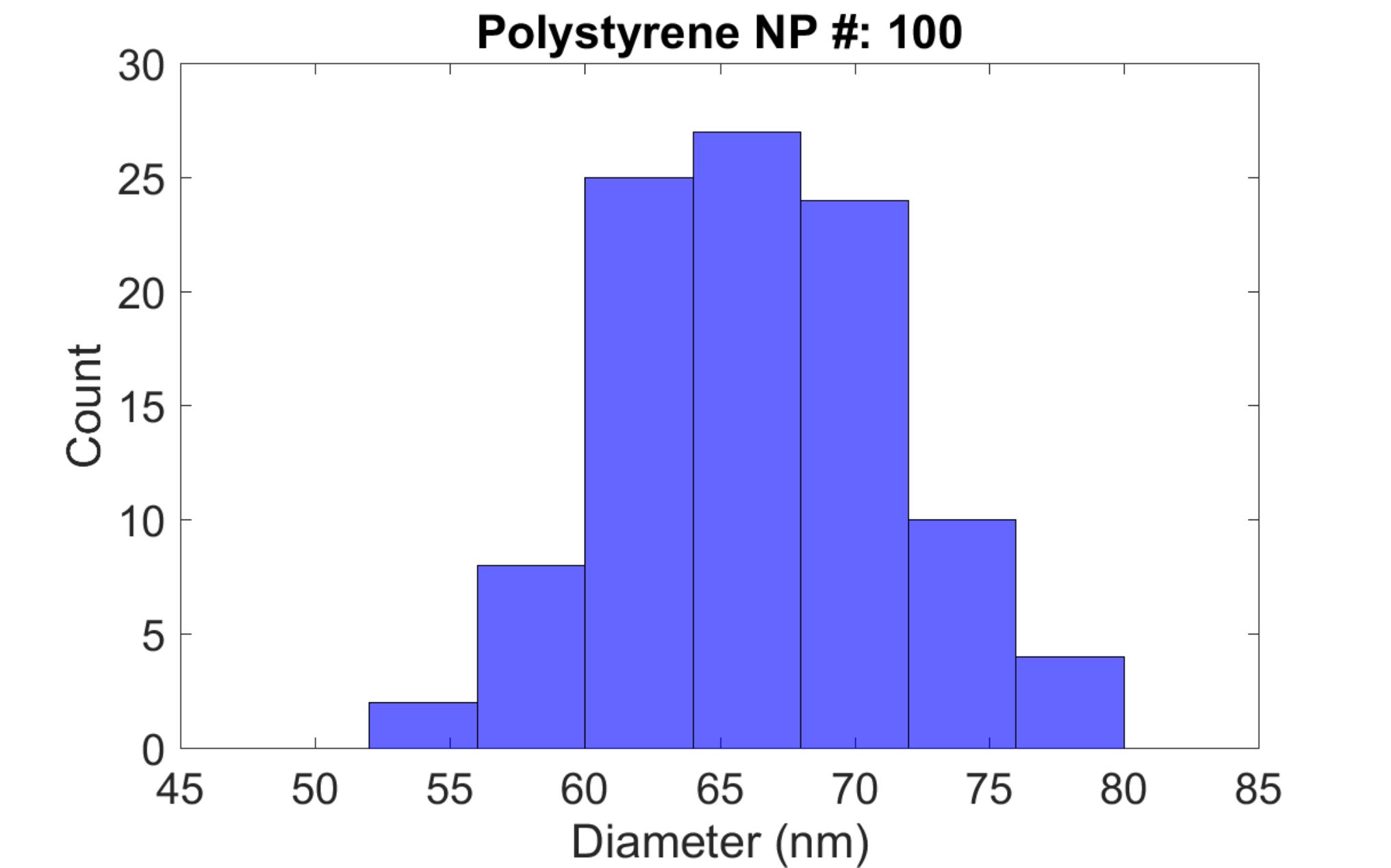}
\captionsetup{justification=centering}
\captionsetup{oneside,margin={.5\linewidth,0cm}}%
\caption{}
\label{fig:fig2e}
\end{subfigure}
\caption{(a) Experimentally obtained defocus curves of gold nanospheres upon supervised learning based type classification. (b) Image of gold and polystyrene nanospheres circled in yellow and blue, respectively. (c) Experimentally obtained defocus curves of polystyrene nanospheres upon supervised learning based type classification. (d) \& (e) Nanoparticle size histograms of the type-classified gold and polystyrene nanospheres, respectively.}
\label{fig:fig2}
\end{figure*}

\section{Classification algorithm}
Support vector machine (SVM), a supervised learning algorithm, can realize the hyperplane with the largest margin that optimizes the separation of a given labeled training data set of $(\boldsymbol{x_i},y_i)$, with $\boldsymbol{x_i}$ being a vector with N features and $y_i$ being its label, resulting in the classification function, $f(\boldsymbol{x})=y $ \cite{SVMcancertissue}. The hyperplane is in the N-dimensional vector space (i.e., $R^N$) of N features, which is the total number of the data points of the vector $\boldsymbol{x_i}$. Moreover, the hyperplane in SVM classifier realizes the linear separation of the corresponding labels of the input vectors. In this context, we employ SVM on nanoparticle classification in terms of type and size distribution. \\
As discussed previously, in wide-field common-path interferometric microscopy, particles exhibit unique defocus signatures that help us realize their size and type. This study relies on the characteristics of those defocus signatures to propose a classification scheme in terms of nanoparticle type and size by learning from the simulations, making use of SVM classifier. The training data set contains 41 features corresponding to the interferometric nanoparticle response at 41 defocus points whose range is determined by the depth of field of the optical system. The simulations span the following parameter space for gold and polystyrene nanoparticles: $15$ nm $\leq r \leq 75$ nm,  $520$ nm $\leq \lambda \leq 530$ nm, where $r$ is the nanoparticle radius, and $\lambda$ is the illumination wavelength. \\
For sake of convenience, we use the built-in \textit{Matlab} SVM classification functions, \textit{fitcsvm} and \textit{fitcecoc}, which can support binary and multiple classes, respectively. The SVM classifier is set to a nonlinear kernel function (e.g., \textit{Gaussian}) to handle the complexity due to the nonlinear features of the vectors \cite{SVMgeneselection}, providing more accurate predictions with the test data set on the trained classifier. The supervised learning for particle type determination considers only the dielectric characteristics of the nanoparticle as a class label, whereas the supervised learning for particle size determination considers only the radius ($r$) as a class label for a given type. We compartmentalize the classification into two steps: we first classify the type, and then proceed with size determination for each population. This two-step approach provides more accuracy as opposed to training and classifying the type and the size in one step.

\subsection{Particle type classification}
In common-path interferometric microscopy, the resonant particles (e.g., gold nanoparticles) differ from the low-index ones (e.g., polystyrene nanoparticles) by exhibiting an eqaully strong positive and negative peaks in their defocus signals, owing to their negative dielectric characteristics \cite{AvciPhysicalModel}. This nanoparticle signature difference between resonant gold and non-resonant polystyrene nanoparticles can be observed from Figs. \ref{fig:fig2a} \& \ref{fig:fig2c}. This phenomenon, in turn, renders resonant nanoparticles distinguishable from the low-index nanoparticles in common-path interferometric microscopy. Hence, we are able to identify the particle types using the built-in Matlab SVM function (\textit{fitcsvm}), associated with binary-class model classification upon training with the simulated defocus curves that are single-labeled according to the particle type.

\subsection{Particle size classification}
For a given particle type, size distribution of the particle can be determined owing to its radius dependent defocus signal. The amplitude of the defocus signal is size dependent such that it scales with the particle polarizability which is a function of particle volume. For instance, the peak-to-peak normalized intensities for polystyrene spheres of radius 25, 30 and 35 nm are 2.5\%, 3.4\%, and 6.5\%, respectively. Thus, SVM can realize the size distribution of a given particle population from the size-dependent peak-to-peak intensity value of the defocus signal. The built-in Matlab SVM function, \textit{fitcecoc}, is capable of multiclass model training. We use this function to train the classifier with the simulated data set for a particular type of nanoparticles of various sizes ($15$ nm $\leq r \leq 75$ nm) with the labels corresponding to the particle size.

\section{Experimental results}
In this section, we experimentally verify our nanoparticle classification algorithm. The experiment uses a sample that consists of gold nanospheres of 29 nm nominal radius and polystyrene nanospheres of 37 nm nominal radius spin-coated on a 100 nm SiO$_2$/Si sensor chip. The experimental defocus scan of the sample is taken using the wide-field  interferometric microscopy setup detailed in \cite{Avci:17}, and the data is processed with a custom Matlab code that performs difference of Gaussian spatial filtering to find nanoparticles in the image \cite{JSTQE}. It is important to note that the code finds the locations of the diffraction-limited nanoparticles in the images regardless of their type and signal strength. Following this initial step of nanoparticle detection in the images; for each nanoparticle, we obtain its defocus response. Each defocus data is then first processed with the classification algorithm for the type determination. Fig. \ref{fig:fig2b} shows a sample image where gold and polystyrene nanospheres are distinguished from one another in the same field of view using this classification algorithm. Notice that the type classification works well due to the unique defocus curves that gold and polystyrene nanospheres exhibit as shown in Figs. \ref{fig:fig2a} \& \ref{fig:fig2c}. That is to say, resonant metallic nanoparticles exhibit an interferometric defocus signal that has a strong positive and negative peaks within an approximately micron defocus range, whereas the non-resonant low-index nanoparticles only exhibit a strong positive peak, as illustrated in Fig. \ref{fig:fig1b}.
Second part of the classification algorithm focuses on the size determination of the particles taking into account their previously determined types. Since the normalized signal strength in the defocus data is size dependent, different sized nanoparticles of the same type can easily be distinguished among themselves. Therefore, by dividing the classification algorithm in two steps, not only do we simplify the classification problem that is normally nonlinear, but also ensure high accuracy. We experimentally verify the second step of this algorithm by further processing the previously detected and type-classified gold and polystyrene nanoparticle signals. We obtain the size histograms of the detected nanoparticles as presented in Figs. \ref{fig:fig2d} \& \ref{fig:fig2e}, which  are in agreement with their nominal sizes provided by their manufacturers.

\section{Discussion and conclusion}
In this study, we have demonstrated the nanoparticle classification in a wide-field interferometric microscopy scheme by combining its powerful nanoparticle imaging capability with a model-based supervised learning algorithm. The proposed classification method accomplishes high accuracy as experimentally verified in Fig. \ref{fig:fig2}. It is imperative to note that, the nanoparticle signal in common-path interferometric microscopy depends on various parameters as examined in \cite{AvciPhysicalModel}. The multivariable dependency of the interferometric signal poses a challenging problem to decipher its constituents. However, this multi-parameter-dependent nanoparticle signal also opens up possibilities in terms of target particle analysis, following its detection. Certain priori information/assumption is therefore necessary not only to simplify this inverse classification problem but also to maintain accuracy. One of the limitations imposed by this particle classification scheme is its size range and shape. 

In the scope of this study, we only assume particles of spherical shape, though, the polarizability calculations can be extended to incorporate the classification of prolate particles. The size constraint, on the other hand, stems from the fact that the dipole approximation that is used in the physical model starts to break down as the particle size gets closer to the illumination wavelength \--- i.e., the electric dipole model to represent the scattering from nanoparticles holds true only when the size is much smaller than the wavelength. It is imperative to note that the dielectric characteristics of the particles are predetermined when carrying out the size classifications. Typically, the biological particles that are of great interest in biomedical studies exhibit similar non-resonant dielectric characteristics as polystyrene nanoparticles.

In conclusion, we have successfully demonstrated a nanoparticle classification scheme in two steps: first determining the particle type and then realizing its size distribution. We have shown the utility of our model-based classification method in differentiating resonant gold nanoparticles from non-resonant polystyrene nanoparticles, as well as, in sizing of the type-classified gold and polystyrene particles whose nominal radiuses are given as 29 nm and 37 nm, respectively. Our classification model has a potential impact in nanoparticle detection using interferometric microscopy, where it allows for simultaneous use of labeled (e.g., with gold nanoparticles) and label-free detection modalities, as well as sensitivity improvements by accurately eliminating the count of nonspecifically-bound particles, and further information about target particles in terms of their type and size.

\section{Acknowledgments}
O. A. gratefully acknowledges support from I/UCRC Center for Biophotonic Sensors and Systems (CBSS).

\bibliographystyle{ieeetr}
\nocite{*}

\bibliography{bib_file}

\begin{thebibliography}{10}

\bibitem{AvciReview}
O.~Avci, N.~L. \"{U}nl\"{u}, A.~Y. Ozkumur, and M.~S. \"{U}nl\"{u},
  ``Interferometric reflectance imaging sensor (IRIS) \--- a platform
  technology for multiplexed diagnostics and digital detection,'' {\em
  Sensors}, vol.~15, no.~7, p.~17649, 2015.

\bibitem{AvciPhysicalModel}
O.~Avci, R.~Adato, A.~Y. Ozkumur, and M.~S. \"{U}nl\"{u}, ``Physical modeling
  of interference enhanced imaging and characterization of single
  nanoparticles,'' {\em Opt. Express}, vol.~24, pp.~6094--6114, Mar 2016.

\bibitem{Avci:17}
O.~Avci, M.~I. Campana, C.~Yurdakul, and M.~S. \"{U}nl\"{u}, ``Pupil function
  engineering for enhanced nanoparticle visibility in wide-field
  interferometric microscopy,'' {\em Optica}, vol.~4, pp.~247--254, Feb 2017.

\bibitem{NovotnyBook}
L.~Novotny and B.~Hecht, {\em Principles of nano-optics}.
\newblock Cambridge university press, 2012.

\bibitem{Daaboul:14}
G.~G. Daaboul, C.~A. Lopez, J.~Chinnala, B.~B. Goldberg, J.~H. Connor, and
  M.~S. Ünlü, ``Digital sensing and sizing of vesicular stomatitis virus
  pseudotypes in complex media: A model for ebola and marburg detection,'' {\em
  ACS Nano}, vol.~8, no.~6, pp.~6047--6055, 2014.
\newblock PMID: 24840765.

\bibitem{Scherr:16}
S.~M. Scherr, G.~G. Daaboul, J.~Trueb, D.~Sevenler, H.~Fawcett, B.~Goldberg,
  J.~H. Connor, and M.~S. \"{U}nl\"{u}, ``Real-time capture and visualization
  of individual viruses in complex media,'' {\em ACS Nano}, vol.~10, no.~2,
  pp.~2827--2833, 2016.
\newblock PMID: 26760677.

\bibitem{JSTQE}
J.~Trueb, O.~Avci, D.~Sevenler, J.~Connor, and M.~S. \"{U}nl\"{u}, ``Robust
  visualization and discrimination of nanoparticles by interferometric
  imaging,'' {\em IEEE Journal of Selected Topics in Quantum Electronics},
  vol.~PP, no.~99, pp.~1--1, 2016.

\bibitem{Unlu:2015}
M.~S. \"{U}nl\"{u}, ``Digital detection of nanoparticles: Viral diagnostics and
  multiplexed protein and nucleic acid assays,'' {\em MRS Proceedings},
  vol.~1720, Jan 2015.

\bibitem{imageclassification}
O.~Chapelle, P.~Haffner, and V.~N. Vapnik, ``Support vector machines for
  histogram-based image classification,'' {\em IEEE transactions on Neural
  Networks}, vol.~10, no.~5, pp.~1055--1064, 1999.

\bibitem{facerecognition}
B.~Heisele, P.~Ho, and T.~Poggio, ``Face recognition with support vector
  machines: global versus component-based approach,'' in {\em Proceedings
  Eighth IEEE International Conference on Computer Vision. ICCV 2001}, vol.~2,
  pp.~688--694 vol.2, 2001.

\bibitem{humandetection}
N.~Dalal and B.~Triggs, ``Histograms of oriented gradients for human
  detection,'' in {\em 2005 IEEE Computer Society Conference on Computer Vision
  and Pattern Recognition (CVPR'05)}, vol.~1, pp.~886--893 vol. 1, June 2005.

\bibitem{EMG}
A.~Subasi, ``Classification of emg signals using pso optimized svm for
  diagnosis of neuromuscular disorders,'' {\em Computers in biology and
  medicine}, vol.~43, no.~5, pp.~576--586, 2013.

\bibitem{SVMcancertissue}
T.~S. Furey, N.~Cristianini, N.~Duffy, D.~W. Bednarski, M.~Schummer, and
  D.~Haussler, ``Support vector machine classification and validation of cancer
  tissue samples using microarray expression data,'' {\em Bioinformatics},
  vol.~16, no.~10, pp.~906--914, 2000.

\bibitem{SVMgeneselection}
I.~Guyon, J.~Weston, S.~Barnhill, and V.~Vapnik, ``Gene selection for cancer
  classification using support vector machines,'' {\em Machine learning},
  vol.~46, no.~1-3, pp.~389--422, 2002.

\end{thebibliography}

\end{document}